# Computer Arithmetic Preserving Hamming Distance of Operands in Operation Result

Shlomi Dolev
Department of Computer Science,
Ben-Gurion University, Beer-Sheva, Israel
E-mail: dolev@cs.bgu.ac.il

Sergey Frenkel
Institute of Informatics Problems,
Russian Acad of Sc., Moscow, Russia
E-mail: fsergei51@gmail.com

Dan E. Tamir
Department of Computer Science,
Texas State University, San Marcos,
Texas 78666, USA
E-mail: dt19@txstate.edu

*Abstract*— The traditional approach to fault tolerant computing involves replicating computation units and applying a majority vote operation on individual result bits. This approach, however, has several limitations; the most severe is the resource requirement. This paper presents a new method for fault tolerant computing where for a given error rate, $r$, the hamming distance between correct inputs and faulty inputs as well as the hamming distance between a correct result and a faulty result is preserved throughout processing thereby enabling correction of up to $r$ transient faults per computation cycle. The new method is compared and contrasted with current protection methods and its cost / performance is analyzed.

## I. Introduction

Fault tolerance is a fundamental concern in the area of communication, data storage, and reliable computation [1]. Error correction in memory units is an established area of research and development. Recently, however, error correction of soft-errors, in specific soft-error upset (SEU) during processing, became a major topic of research and interest [2]. One of the most important means for error correction is the exploitation of spatial and / or temporal redundancy.

Efficient, error correcting codes (ECC), such as Hamming codes applied in channel coding do not duplicate bits. Instead, they use methods that exploit the Hamming distance between code-words to correct errors. In the case of processing, however, triple modular redundancy (TMR) along with majority vote over replicates of computing units is the most commonly used approach for fault correction [3].

The TMR method is simple. Duplicating units, however, is quite expensive. Moreover, TMR has an inherent problem where the majority vote operation performed at the final stage of computation is not protected. Hence complete fault tolerance cannot be guaranteed. To cope with TMR limitations, transient and SEU ECC based protection techniques have been reported; and presently, ECC is included in various VLSI devices and microprocessors [4]. Nevertheless, ECC in VLSI devices is generally used for protecting data and memory units and is not applied efficiently for protecting the actual computation and processing [4,5,6,7].

Hamming coding, used to correct 1-bit errors, is efficient ECC based protection method. When the probability of more than one error in a data block is high, and in the case of VLSI devices where a single defect can cause multiple output errors, other types of linear codes, such as Bose–Chaudhuri-Hocquenghem (BCH) coding and Reed-Solomon (RS) coding should be considered [8].

In this paper we address the logical level of digital circuits and propose a novel method for using Hamming, BCH, and RS codes for protecting computation results in processing units. Since memory protection has already been extensively researched and developed, we concentrate on providing high reliability with respect to permanent and transient faults for operational blocks such as the arithmetical-logical unit (ALU). We analyze the cost / performance of the new method and compare it with existing methods.

The rest of the paper is organized as follows: Section II describes related work. Section III presents the ALU architecture for the Hamming processor, and describes our method for Op-code protection. Section IV presents the BCH and Reed-Solomon extensions for covering more than one error. Section V analyzes cost, performance, and practicability issues. Finally, Section VI includes conclusion and proposals for future research.

## II. RELATED WORK

Generally, the circuits used to protect VLSI and micro-processors are based on TMR, ECC, or a combination of the two approaches [9]. Nevertheless, the traditional use of ECC concentrates on memory elements and provides protection for the outputs of individual functions through their coding [5,6,7,9,10]. Jasinski et al. discuss the utility of Hamming coding for error protection in arithmetic blocks and assert that Hamming coding should be considered in this scope [7].

Traditional TMR techniques assume that the combinational structure of a circuit is not subject to bit flips due to transient errors. In modern nanotechnology designs, however, higher clock frequencies increase the probability of capturing glitches [11,12]. Hence, as noted by several authors; in the context of protection of sequential elements, combinational logic is the dominant source of the transient faults [13].

Recently, new and advanced approaches for protecting combinational logic from transient faults have been devised. These techniques add time redundancy with double edge triggered registers to the space redundancy of the traditional TMR [13]. Nevertheless, two important limitations of traditional and advanced TMR are the degree of redundancy and the possibility that a SEU hits the voting gate before a primary output does.

Hamming coding techniques are considered as a viable alternative to TMR because they require low-complexity decoding circuits; and are highly efficient for correcting a single error. Nevertheless, these techniques are primarily used for memory and register file protection. Lima et al. consider a microprocessor Data Path protection schema, but they address only the protection of the ALU registers' output [10]. On the other hand, the architecture suggested by Hass et al. assumes an encoder and a decoder for each operation [11]. This architecture, however, cannot protect the encoding and decoding circuits simultaneously. Finally, the decoding circuit that performs the final error correction in the system proposed by Benschop is not completely protected [6]; although errors in its operation can be detected through the introduction of self checking unit in the proposed circuits. For this reason the decoding circuit must be as small as possible compared to the circuit to be protected.

Numerous publications explore trade-offs between TMR and Hamming codes for micro-architecture level soft error mitigation solutions. For example, according to Nelson, TMR is the best solution if the latency of the register file is most important [3]. In addition, Nelson shows that for memory-rich systems such as microprocessors, the Hamming coding has smaller area overhead (less than half) in comparison with TMR [3,10].

### III. HAMMING DISTANCE PRESERVING ALU

Let $x = [d_x, h_x]$ and $y = [d_y, h_y]$ be two Hamming-encoded operands represented in binary format that includes the operand's bit vectors $d_x$ ($d_y$) along with the Hamming parity bit vectors $h_x$ ($h_y$). Further assume that both $x$ and $y$ are correctable. That is, the error correction functions $C(x')$ ($C(y')$), where $x'$ ($y'$) are versions of $x$ ($y$) that include up to $r$ errors for $x$ ($y$) respectively. The term $r$ represents the correctable distance, of the Hamming code used, that can produce the correct bit vector. We apply a micro-operation $op$, to $x$ and $y$; and the goal is to find a way to perform this operation so that the result is the correctable Hamming code of $C(x)$ $op$ $C(y)$. In other words, let $C(t') = t$ and let $x$ $op$ $y = z$, then it is desired that despite of any faulty gates or faulty inputs that sum-up to less than $r$ errors $C(x)$ $op$ $C(y) = C(z)$.

We present arithmetic to compute the operation $C(x)$ $op$ $C(y)$, while preserving the correct-ability of the Hamming encoded result. If the Hamming distance of $x$ ($y$) from $C(x)$, ($C(y)$) is $d_x$ ($d_y$) and the maximal correctable distance is $d_c$ then our implementation can withstand $d_c - (d_x + d_y)$ faulty gates [4]. We consider both logical and arithmetical operations, namely: bit-wise XOR, AND, OR, NOT, NAND, and NOR as well as addition and subtraction. The examples use a (7,4) Hamming code. In addition, we assume that at most one error can occur (anywhere) in the logic implementing the operation.

Bitwise *XOR* and bitwise *NOT* are the most simple operations in this schema since the parity bits of the Hamming code are based on *XOR* (*NOT*) computations. A circuit that consists of $d + h$ *XOR* (*NOT*) gates, where $d$ is the number of data bits and $h$ is the number of parity bits, can be used to implement the Hamming distance preservation requirement. Each of these gates computes one bit of the result $z = [d_z, h_z]$. Figure 1, shows an example of a (7,4) Hamming encoded *XOR* block. The example corresponds to the assumption of a single transient error. A bitwise *NOT* can be implemented using *NOT* gates or by applying bitwise *XOR* to the operands $x$ and $\bar{1} = [111 \dots 1]$.

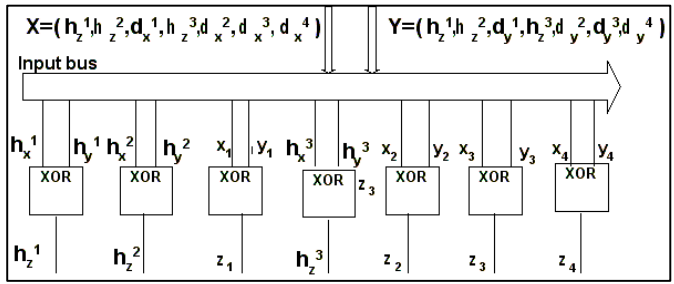

Fig. 1. Bitwise Hamming Distance Preserving *XOR* Block

As the figure shows, the two inputs for such a *XOR* block produce the corresponding bits of the output $z$. The independent computation of bits, in particular the Hamming redundant bits, preserves the location of erroneous bits, which in turn results in correctable $z$.

The implementation of the bitwise operations of *AND*, *OR*, *NOT*, *NAND*, and *NOR* share the same principle. We consider only the *NAND* operation. Figure 2 illustrates the implementation of a bit-wise *NAND* Hamming distance preserving block. The circuit uses a "building-block" that performs 4-bit bitwise *NAND*. *One* of these blocks is used to produce the data bits of $z$ ($d_z$). The other 4-bit bitwise *NAND* blocks generate the parity bits ($h_z^i$). Each of these blocks accepts $C(x)$ and $C(y)$ as inputs, generates data bits of $z$, and feeds them into a parity generator (the *XOR* gates*)*. The end-result is a 7-bit representation of the correctable version of $z$.

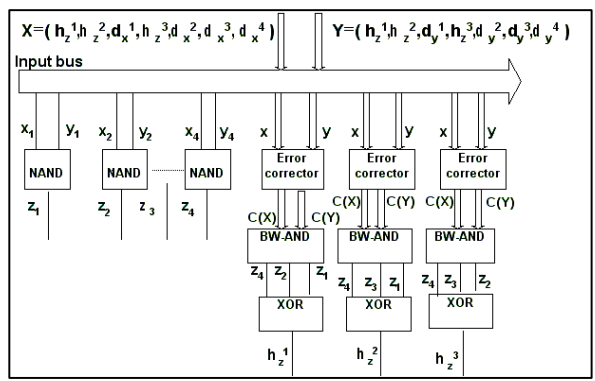

Fig. 2. Bitwise Hamming Distance Preserving *NAND* Block

Under this implementation, assuming a Hamming distance of three and at most one error, if there is an error in $d_z$ and no error in the part of the circuit that computes $h_z^i$, then $h_z$ is pointing to the incorrect output bit of $d_z$. Otherwise one of the bits of $h_z$ is the only corrupted output; hence it is correctable.

Since each of the parity bits requires only three bits of $d_z$, the circuit depicted in figure 2 can be simplified. Moreover, the core of the circuit can be designed "from scratch" using two level logic with 8 inputs ($d_x$ and $d_y$) and 7 outputs $z = [d_z, h_z]$. Furthermore, since *NAND* is a universal (functionally complete) operation, then the circuit of figure 2 along with an actual

Hamming "corrector" circuit can be used as a building block for any fault tolerant 2-level logic implementation of combinatorial circuits.

The principle that drives the implementation of the Bitwise *NAND* Hamming block can be generalized to several other combinatorial units such as a full adder. One set of "building blocks" is used to compute individual bits of the result bits $d_z$ and another set of building-blocks is used to generate individual bits of $h_z$.

Figure 3, shows an implementation of a 4-bit adder. In the figure, the building-block is a 4-bit full adder. A set of 4-bit full adders depicted on the left side of the figure is used to compute individual bits of the result bits $d_z$ and another set of full-adders depicted in the right side of the figure generates the bits $h_z^i$. As in the case of the bitwise *NAND*, the implementation presented in figure 3 can be significantly simplified. For example, rather than duplicating full adders, only the part of the adder that is essential to generate the relevant bits of $d_z$ can be used. In this sense, the implementation presented in the figure should be considered as an illustration of the concept rather than as an optimized implementation. Moreover, the principle can be extended to include input and output carry. Alternatively it can be used with building blocks that consist of a carry look-ahead 2's complement adder / subtractor.

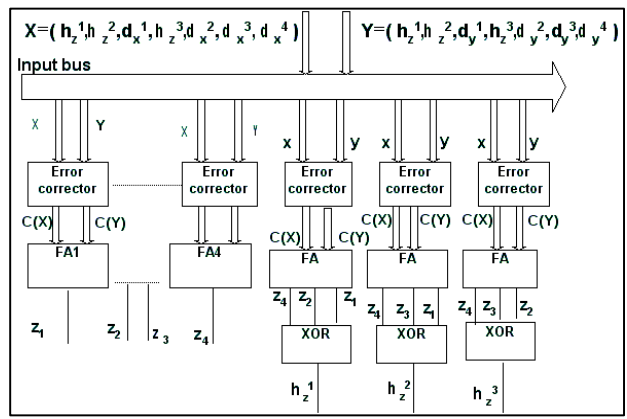

Fig. 3. The architecture of a Hamming Full Adder

An important problem that requires attention is the protection of enable signals in arithmetic blocks [7]. Indeed, the approach proposed in this paper can be used to provide this feature. For example consider a two bit op-code with bits $\{c_1, c_2\}$. In order to protect these enable signals ($\{c_1, c_2\}$) we produce 2 parity bits $\{h_z^i, h_z^e\}$. Now we can use the principle depicted in figure 2, and produce a circuit with two main blocks one block is similar in principle to the blocks depicted in the left part of figures 2. It generates the individual control bits $\{c_1, c_2\}$. The second block which is similar to the right part of the figure produces the parity bits $\{h_z^i, h_z^e\}$. Additional details on the enable protection are included in [14].

### IV. Multi-bits Error Correction

The probability of multiple-bit errors in large and complex circuits can be relatively high. This might necessitate multiple-bit error correction units such as BCH and RS coding based error correction units. Nevertheless, the principle of separating the computations of the data bits from the computation of the error correction bits proposed in this paper can still be exploited; enabling the use of other linear systematic codes, such as, BCH codes in our methodology [8].

In order to detect and correct multiple errors, we need to extend the length of the syndrome vector of the Hamming code in terms of BCH codes, which can handle randomly located errors in a data stream. The only requirement with respect to possible multiple error classes, is that errors are correctable on the operational block output, i.e., the number of error is less than $(D - 1)/2$, where $D$ is the Hamming distance of the code. Hence, we can use an architecture that is analogous to the one used for the Hamming code. The circuit for each of the operations considered includes the BCH error correctors corresponding to the coding used; and the required parity bits are computed by a corresponding BCH circuit.

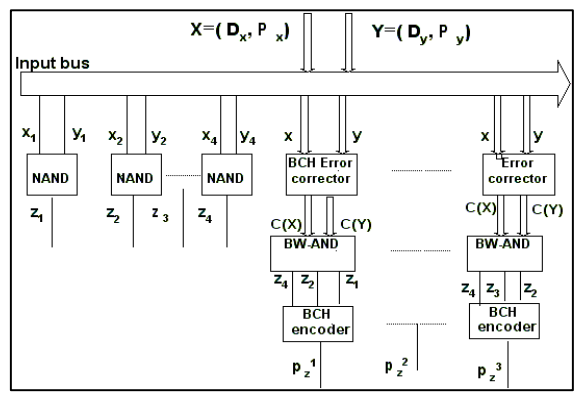

Fig. 4. A $d$-bits, BCH based, Hamming bitwise *NAND* Block.

Figure 4 demonstrates a circuit providing $(n - k)/2$ symbols error correctable bitwise *NAND* computation, where $2t = n - k$ parity bits enable correcting up to $t$ symbols that contain errors in a codeword of length $n$ with $k$ data bit. The vector of data symbols results $Z = (z_1, z_2, ... z_d)$ is the data bits and parity bits output of the bit-wise operations,

### V. Cost, Performance, and Practicality Issues

Perfect Hamming coding, that is, codes designed for 1-bit errors, which satisfy the relation $2^{n-k} = n$, require a circuit with size that is a logarithmic function of the number of code bits [15]. Nevertheless, BCH codes, in particular RS codes, are widely implemented because they can correct multi-bit errors and are "almost perfect" in the sense that the redundant bits added by the encoder is the minimum for any level of error correction [16,17]. Thus, we may consider this growth also as an "almost logarithmic" growth. Hence, the underlying hardware redundancy of our method is a logarithmic function of the error rate ($r$). In addition, it is a logarithmic function of the number of ALU bits (since the number of parity bits is a logarithmic function of the number of data bits). In contrast, a TMR based implementation of an n-bit ALU requires a redundancy factor of $(2r + 1) \times n$. Thus it is a linear function of the error rate and the number of bits.

A drawback of our method, however, is the encoding overhead. The encoding requires bitwise XOR of two symbols, and multiplication of two elements of two polynomials, representing the operands. Nevertheless, these operations can be implemented efficiently and economically using an AND-XOR network, as explained in [18,19,20]. Thus, for large values of $r$ and $n$ the two methods require about the same resources. Hence, it is expected that for a moderate to large values of $r$ and $n$ our method will require less redundancy / overhead than TMR.

The last TMR stage, the majority vote, is not protected. The same applies to the last stage of our method, the Hamming corrector. Thus, the overall probability of error in the two methods is not 0 and total protection is not guaranteed. Nevertheless, there is a significant difference between the two approaches. The TMR of an $n$ bit ALU requires $n$ unprotected majority vote units, while our method requires only one (yet somewhat more complicated) unprotected Hamming corrector unit. Consequently, our method provides a more resilient and cost effective error protection. Moreover, for a large number of errors in combinational circuits, traditional TMR techniques are limited in their error correction-ability [21,22].

It might be interesting to compare our approach to error detection techniques. One of the most popular detection methods is the Berger check prediction (BCP) [23]. The BCP check bit length *is* given by $k = log_2(n + 1)$, where $n$ is the number of bits in the original data word [23]. For example, a 32-bit ALU would require a 6-bit BCP Symbol, while our method would require a 7-bit symbol, but the BCP based circuit cannot correct the errors. Furthermore, the BCP algorithm can only detect unidirectional errors, yet a single fault in a combinational circuit can lead to bidirectional errors and require a BCP unit for each direction. Hence, our method can be less expensive than error detection methods while providing error correction.

## VI. CONCLUSIONS AND FUTURE RESEARCH

This is the first step in incorporating linear ECC in the arithmetical and logical processing phase. In the traditional fault tolerant computing the reliability of a computation is a linear function of the error rate $r$ and the number of bits in a computational unit. In contrast, we suggest the same or better reliability with an amount of resources that is proportional to the logarithm of the error rate and the number of bits. Hence, we expect a better cost performance for moderate-to-large (in terms of number of bits) computational units. In addition, our method is less pruned to errors in the last and unprotected stage. Furthermore, we are proposing architecture for a universal tolerant logic. Given all of these potential benefits, our method is very promising.

Nevertheless, additional research for quantifying the overhead and exploring trade-offs between reliability and size-performance-power is due and might require investigating the following issues:

- Taking into account the sequential logic, which affect the result of the operations, that is influenced by errors in the combinational parts,

- Extending the proposed method to sequential logic (e.g., by using the protected NAND gates with feedback),

- Taking into account the relative area of different units such as the Datapath and the memory [24].